\documentclass{aa}
\usepackage{txfonts}
\usepackage{natbib}
\usepackage{graphicx}
\usepackage{multirow}

\bibpunct[, ]{(}{)}{;}{a}{}{,}

\def\gequ{$\gamma$~Equ}
\def\lib{$33$~Lib}
\def\bcrb{$\beta$~CrB}
\def\astron{\textsc{Astron}}

\def\llm{{\sc LLmodels}}
\def\atl{{\sc ATLAS9}}
\def\width9{{\sc WIDTH9}}

\def\logg{\log(g)}
\def\teff{T_{\rm eff}}

\def\bs{\langle B_{\mathrm{s}} \rangle}

\def\halpha{\mathrm{H}\alpha}
\def\hbeta{\mathrm{H}\beta}
\def\hgamma{\mathrm{H}\gamma}
\def\ddafit{{\sc DDAFit}}
\def\synth3{{\sc Synth3}}

\def\synthmag{{\sc Synthmag}}

\def\Lsun{L_{\odot}}
\def\Rsun{R_{\odot}}
\def\Msun{M_{\odot}}

\def\ll{\lambda\lambda}

\begin{document}

\title{Fundamental parameters of bright Ap stars from wide-range energy distributions and advanced atmospheric models}
\titlerunning{Fundamental parameters of bright Ap stars}

\author{D. Shulyak\inst{1}  \and T. Ryabchikova\inst{2,3}  \and O. Kochukhov\inst{4}}
\authorrunning{D. Shulyak et al.}

\offprints{D. Shulyak, \\
\email{denis.shulyak@gmail.com}}
\institute{
Institute of Astrophysics, Georg-August University, Friedrich-Hund-Platz 1, D-37077 G\"ottingen, Germany \and
Institute of Astronomy, Russian Academy of Science, Pyatnitskaya 48, 119017 Moscow, Russia \and
Universit\"at Wien, Institut f\"ur Astronomie, T\"urkenschanzstra{\ss}e 17, 1180 Wien, Austria \and
Department of Physics and Astronomy, Uppsala University, Box 516, 751 20, Uppsala, Sweden
}

\date{Received / Accepted}

\abstract
{}
{As a well-established procedure for the vast majority of normal main-sequence stars,
determination of atmospheric and stellar parameters
turns to be a challenging process in case of magnetic chemically peculiar stars. Inhomogeneous distribution of chemical elements and
strong magnetic fields make most of the standard photometric and spectroscopic calibrations inapplicable for this class
of stars. In this work we make use of available observed energy distributions calibrated to absolute units, stellar parallaxes, high-resolution
spectroscopic observations, and advanced stellar atmosphere models to derive parameters of three bright Ap stars: \lib, \gequ, and \bcrb.}
{Model atmospheres and fluxes were computed with the \llm\ code. \synth3\ and \synthmag\ codes were used to compute
profiles of individual spectral lines involved in abundance analysis.}
{For each of the stars, we construct a self-consistent atmospheric models assuming normal and depleted helium compositions
and derive empirically stratification profiles of certain elements. The effective temperatures and surface gravities are
found from the simultaneous fit to spectroscopic, photometric, and spectrophotometric observations calibrated to absolute
units. We show that using advanced model atmospheres and accurate stellar parallaxes allows one to derive stellar radii
with high accuracy, and these are consistent with those obtained from independent but more complicated interferometric observations.}
{}

\keywords{stars: chemically peculiar -- stars: atmospheres -- stars: abundances -- stars: fundamental parameters -- stars: individual: HD~137909, HD~137949, HD~201601}

\maketitle

\section{Introduction}

Accurate knowledge of fundamental stellar (mass, radius, luminosity, metallicity, age) 
and atmospheric (effective temperature, surface gravity, atmospheric abundances) 
parameters is essential for many fields of modern astrophysics.
Stellar masses and abundances determine the whole chain of stellar evolution stages of a single star, and obtaining stellar
luminosities and radii by means of observed data allows one to determine ages of stars, as well as to derive other
parameters of interest, such as atmospheric chemistry from spectroscopic observations, to characterize extrasolar planets from transit data, 
to carry out asteroseismic modeling, and much more.

Determining stellar parameters is a delicate process, even in the case of normal stars, because it relies on
theoretical evolutionary and atmospheric models. The former requires knowing metallicities and masses, the latter
also needs effective temperatures and gravities as input.
These are usually obtained from spectroscopic and photometric data employing well-developed calibration schemes.
For chemically peculiar (CP) stars, however, things get much more complicated mainly because of the abnormal chemical composition
of their atmospheres \citep[see, e.g.,][]{2004A&A...423..705R}. This makes the spectra and energy distributions of CP stars look very different
from their normal
analogs, and therefore most of the photometric calibrations simply fail to work for CP stars. Furthermore, the
strong surface magnetic fields detected in many CP stars  additionally complicate the analysis of their spectra
and determination of atmospheric parameters.

Since all parameters are linked, the accuracy in derived parameters, such as $\teff$ and $\logg$, must be 
as high as possible.
This is why the full set of stellar and atmospheric parameters can ideally only be derived using an iterative procedure where
each parameter is adjusted in such a way as to fit all available observed data. Then, if needed, the other parameters are recalculated
accordingly. For instance, deriving $\teff$ and $\logg$ from photometric data and employing relevant calibrations one can carry out
a detailed abundance analysis based on model atmospheres, but if derived abundances turn out to be different 
from those assumed in the construction of the model atmospheres then the latter must be recalculated with the appropriate chemistry, 
and the whole process should be repeated again.
Because of the lack of appropriate calibrations and the wide diversity of abundance patterns, this is the only robust scheme 
for analyzing CP stars \citep[see, e.g.,][]{2009A&A...499..851K,2009A&A...499..879S,2011MNRAS.417..444P}.

Recently, interferometry has been successfully applied to CP stars resulting in a first estimate of the radii
of three stars: $\alpha$~Cir \citep[HD~128898,][]{2008MNRAS.386.2039B}, \bcrb\  \citep[HD~137909,][]{2010A&A...512A..55B}, and
\gequ\ \citep[HD~201601,][]{2011A&A...526A..89P}. No doubt interferometry is a powerful technique with the strong
advantage of providing model-independent estimates of stellar sizes, but its current world-wide facilities, unfortunately, are limited
only to bright stars.

In this work we apply a self-consistent atmospheric analysis to three well-known magnetic CP stars: \bcrb~(HD~137909), 
\gequ~(HD~201601), and \lib~(HD~137949).
Collecting available photometric, spectrophotometric, and spectroscopic observations we perform detailed
abundance and stratification analysis of a number of chemical elements in the atmosphere of each star.
The effective temperatures and gravities are found iteratively by comparing theoretical fits
to all available observables. 
Using fluxes calculated consistently with model atmospheres, available stellar parallaxes, and
observed fluxes calibrated to absolute units, we derive stellar radii and compare them with
interferometric results.

\section{Observations}

Our investigation is based mainly on the low-resolution spectroscopic observations
calibrated to absolute units and 
obtained by \textbf{S}pace \textbf{T}elescope \textbf{I}maging \textbf{S}pectrograph (STIS)\footnote{http://www.stsci.edu/hst/stis}
mounted at the Hubble Space Telescope.
STIS provides wide wavelength coverage ($\ll1700-10\,200$) that is excellent for  accurate analysis of the energy distributions
of early type stars where both UV and visual fluxes are highly important. Among our targets, only \lib\ was not observed with STIS. Nevertheless,
medium-band spectro-photometric observations by \citet{1989A&AS...81..221A} are available for all objects, as are UV fluxes from the
International Ultraviolet Explorer (IUE)\footnote{http://archive.stsci.edu/iue/}.
For \lib\ and \bcrb, additional wide-band UV fluxes were taken 
from the TD1 space mission \citep{1973MNRAS.163..291B,1978csuf.book.....T}. For \bcrb\ we also made use of the spectro-photometric measurements 
of \citet{1996BaltA...5..603A},
\citet{1976ApJS...32....7B}, \citet{1995OAP.....8....3K}, and UV low-resolution fluxes from the \astron\ space satellite \citep{1986INTSA..31..198B}.

We note that CP stars are known to demonstrate photometric variability, most of which is caused
by the inhomogeneous surface distribution of chemical elements. However, as can be seen from e.g.
\citet{2010A&A...524A..66S} and \citet{2012A&A...537A..14K},
the amplitude of this variability is rather small to influence the derived effective temperatures much.
For instance, using Adelman's spectrophotometric catalog, we find that the relative characteristic correction to the
effective temperatures due to flux variability is only a few ten K's for all three target stars.
Because of this, and also because the datasets listed above are not homogeneous in the sense of observing times and number of
observations performed, we used mean fluxes when possible. For instance,
only phase-averaged fluxes were used from Adelman's catalog and the IUE data archive.


Chemical abundance and stratification analysis was based on high-resolution, high signal-to-noise-ratio spectra obtained with the UVES instrument 
\citep{DOK00} at the ESO VLT. Observations of \lib\ and \bcrb\ were carried out in the context of program 68.D-0254,
and published by, e.g., \citet{2008A&A...480..811R}, while spectra of \gequ\ were extracted from the ESO archive (program 76.D-0169). Spectra of all three stars were
obtained using both dichroic modes, thus covering wavelength intervals from $\lambda 3030$ to $\lambda 10400$ with a few gaps. Spectral resolution is about 
80\,000, and the signal-to-noise ratio at $\lambda$5000 is about 300-400. A detailed description of the data reduction is given by 
\citet{2008A&A...480..811R} (\lib\ and \bcrb) and by \citet{2007MNRAS.377.1579C} (\gequ). 
In addition, for \lib\ and \gequ\ we used the UVES slicer spectra covering the $\ll 4960-6990$ region 
and providing a resolving power of $\lambda/\Delta\lambda\approx115\,000$. 
Time-resolved observation of these stars were obtained in the context 
of ESO programs 72.D-0138 and 077.D-0150 \citep[e.g.][]{KEM06}. 
We used average spectra constructed from the total of 111 and 178 short exposures for \lib\ and \gequ, respectively. 
Reduction of these data followed procedures described by \citet{2007A&A...473..907R}.

In the course of our analysis we found that for the \bcrb\ and \gequ\ the $\hbeta$ line of the
UVES spectra always require systematically different temperatures compared to the one derived from the spectral
energy distribution (SED) and other
hydrogen lines. We attribute this to a problem of continuum normalization and made use of additional
observations from different instruments. In particular, for \bcrb\ we used spectroscopic data obtained at $2.6$-m
telescope mounted at CrAO (Crimean Astrophysical Observatory) \citep{1998AstL...24..516S}.
For \gequ\ we used medium resolution spectra ($R\approx37~000$)
obtained with the echelle spectrograph GIRAFFE, mounted at the $1.9$-m Radcliffe telescope of SAAO 
(South African Astronomical Observatory).
The data reduction of the GIRAFFE spectra was based on Vienna automatic 
pipeline for echelle spectra processing described in \citet{2003IAUS..210P.E49T}.

\section{Basic methods}
\label{sec:methods}

\subsection{Model atmospheres}

Our approach is based on the models and synthetic fluxes 
computed with the \llm\ stellar model atmosphere code \citep{llm}.
The code can treat individual homogeneous and depth-dependent abundances, which is a key
feature in our modeling. In addition, the effect of the magnetic field is fully accounted
via detailed computations of the anomalous Zeeman splitting and polarized radiative
transfer as described in \citet{2006A&A...448.1153K}. In all calculations we adopted magnetic field
vector to lay perpendicular to the normal of the stellar surface \citep[see][for more details]{2006A&A...454..933K}.
This makes \llm\ a powerful tool for the modeling of
the observed characteristics of magnetic CP stars. The model fluxes are computed on a fine
frequency grid with the highest point density at the wavelengths occupying the region of maximum stellar flux.
A total of about $600\,000$ frequency points are usually used, which provides the necessary resolution
for accurate computation of photometric indicies of broad- and narrow-band photometric systems.
Lines opacity coefficient is computed based on atomic parameters extracted from the {\sc VALD} database \citep{vald1,vald2},
which presently includes more than $6.6\times10^7$ atomic lines. Most of them come from the latest
theoretical calculations performed by R. Kurucz\footnote{http://kurucz.harvard.edu}.

Because of their large overabundance in atmospheres of Ap stars, rare-earth elements (REE) play an important
and sometimes even dominating role in the appearance of spectroscopic features \citep{2009A&A...495..297M,2005A&A...441..309M}
and can also affect the SED \citep{2010A&A...520A..88S}. Here we used the same line lists of
REE elements as presented in \citet{2010A&A...520A..88S}.
The studies by \citet{2009A&A...495..297M,2005A&A...441..309M} demonstrate that the line formation 
of {Pr}{} and {Nd}{} can strongly deviate from the local thermodynamic equilibrium (LTE) with
the doubly ionized lines of these elements are unusually strong 
due to the combined effects of stratification of these elements and departures from LTE 
(known among spectroscopists as the ``REE anomaly''). 
However, a precise NLTE analysis is beyond the scope of this paper and currently
could not be directly coupled to a model atmosphere calculation. Therefore, we followed the approach outlined in \citet{2009A&A...499..879S} 
where the authors used a simplified treatment of the REE NLTE opacity by reducing the oscillator strengths for
the singly ionized REE lines by the corresponding LTE abundance difference, while using the abundances derived from second ions 
as input for model atmosphere calculations (or vice versa). 
Applying this line scaling procedure allowed us to mimic the line strengths that correspond to the NLTE ionization 
equilibrium abundance for each REE.

\subsection{Abundance and stratification analysis}

Initial homogeneous abundances for model calculations were taken from the papers by \citet{2004A&A...423..705R} (\lib\ and \bcrb) and by 
\citet{1997A&A...322..234R} for \gequ. These abundances were slightly corrected with synthetic spectral synthesis taking into account rather large
magnetic fields of the program stars. Spectrum synthesis was performed by the magnetic synthesis code {\sc SYNTMAG} \citep[see][]{synthmag07} using
extraction of the atomic parameters and Land\'e-factors from the {\sc VALD} database. 
To make abundance analysis of magnetic stars more efficient, we used the equivalent width method with the newly developed 
code {\sc WidSyn}. This program provides an interface to the {\sc SYNTMAG} code and allows abundances to be derived
using theoretical equivalent widths obtained from the full polarized radiative transfer spectrum synthesis. 
This improved equivalent width method considerably accelerates abundance analysis without sacrificing accuracy.
  
The stratification of chemical elements was modeled by employing the \ddafit\ IDL-based automatic procedure 
that finds chemical abundance gradients from the observed spectra \citep[see][]{synthmag07}. In this routine, the vertical
abundance distribution of an element is described with
the four parameters: chemical abundance in the upper atmosphere,
abundance in deep layers, the vertical position
of abundance step, and the width of the transition region
where chemical abundance changes between the two values.
All four parameters can be modified simultaneously with the least-squares fitting procedure and based on observations
of an unlimited number of spectral regions. This procedure
was successfully employed by a number of element stratification studies 
\citep[e.g.,][]{2005A&A...438..973R,2009A&A...499..851K,2009A&A...499..879S,2011MNRAS.417..444P}.

For {Pr}{} and {Nd}{} in \gequ, we performed an NLTE stratification analysis as 
described in \citet{2005A&A...441..309M,2009A&A...495..297M}  using a
trial-and-error method and the observed equivalent widths of the lines of the first and second ions.
In NLTE calculations magnetic field effects were roughly approximated by pseudomicroturbulent velocity
individual for each line. Such an analysis is applicable only for stars with small
or moderate magnetic fields, e.g. for HD~24712, \gequ\ \citep{2005A&A...441..309M,2009A&A...495..297M}.

\subsection{Determination of fundamental parameters}

We follow the basic steps of the iterative procedure of atmospheric parameters determination
outlined in, e.g., \citet{2009A&A...499..879S}. Briefly, this procedure consists
of repeated steps of stratification and abundance analysis that
provide an input for the calculations of model atmosphere until
atmospheric parameters ($\teff$, $\logg$, etc.) converge and theoretical
observables (energy distribution, profiles of hydrogen
lines, etc.) fit observations. Thus, the derived abundance pattern is consistent with the physical
parameters of the final stellar atmosphere.

We used low-resolution spectroscopic and/or spectro-photometric data calibrated to
absolute units and available stellar parallaxes to derive radii of investigated stars. 
Fitting the full energy spectrum makes it possible to find reliable values for the 
atmospheric $\logg$ by matching the amplitudes of Balmer jump, 
which is very difficult or even impossible to do when only spectroscopic indicators or photometric indices are used.
The fit to observed energy distributions is carried out on each iteration of abundance analysis 
using a grid of model atmospheres computed for a set of $[\teff$,$\logg]$ values. The radii of
the star is found by minimizing the deviations between observed and predicted absolutely
calibrated fluxes.

However, one important uncertainty remains in this analysis. As follows from the computations of atomic diffusion, helium always sinks 
in subphotosheric layers of A-F stars \citep{1979ApJ...234..206M}. The same He depletion is  also found
by the diffusion models of the optically thin layers \citep{2004IAUS..224..193L}, suggesting that the atmospheres of CP stars of spectral
types A-F may well be He-deficient. No suitable diagnostic He lines are available for cool Ap stars to verify this conclusion observationally.
The question of whether helium is indeed removed from the atmosphere is particularly important because
this may noticeably change the molecular weight and T-P stratification of He-weak models in deep layers
where temperatures become high enough for He to contribute to the total opacity coefficient.
Having no clues to the true He stratification, we thus consider an additional case of He-weak atmosphere
with adopted He abundance of [He/H]$=-4$ (decreasing He abundance below this value has only a marginal or no
effect).

\section{Results}
\label{sec:results}

In this section we present the results of the parameter determinations for individual Ap stars, which are
combined in Table~\ref{tab:results}. For completeness, this table also lists our previous results derived
for a few additional CP stars using the same analysis methodology.
We also find that, with a few exceptions, the application of realistic model atmospheres did not result 
in a significant change of the mean element abundance estimates with respect to the studies cited above. 
Therefore, in the following we focus on the analysis of the vertical abundance profiles.

\subsection{\bcrb\ (HD~137909)}

Among the three targets, \bcrb\ is probably the most studied one. A number of independent
spectrophotometric observations of energy distributions make it possible to test predictions of atmospheric models,
as well as estimate uncertainties of derived fundamental parameters in detail.
Furthermore, \bcrb\ is known to be a binary system with orbital elements presented by \citet{1984PAZh...10..293T} and
\citet{1998A&AS..130..223N}.
More recently, \citet{2010A&A...512A..55B} have carried out interferometric observations of the system, which resulted 
in the first model-independent 
determination of the radii of primary (A) and secondary (B) components: $R(A)=2.63\pm0.09\Rsun$, $R(B)=1.56\pm0.07\Rsun$.
However, authors still used scaled-solar models computed with \atl\ model atmosphere code \citep{1992IAUS..149..225K,1993KurCD..13.....K}
to predict bolometric flux of the system and thus to derive effective temperatures for both
components. This was necessary because all modern flux data always contain
the combined light of the two components. This is also true for the STIS observation, which was done with the $0.2$\arcsec\ slit
width on 25~Aug~2003 when the visible separation between system components was $r<0.1$\arcsec.
Effective temperatures were found to be $\teff(A)=7980\pm180K$ and $\teff(B)=6750\pm230$~K, respectively.
The secondary component thus provides an important $\approx18$\%\ contribution to the bolometric flux of the system
and thus has been taken into account in our SED fitting procedure.

\begin{figure*}
\centerline{\includegraphics[width=\hsize]{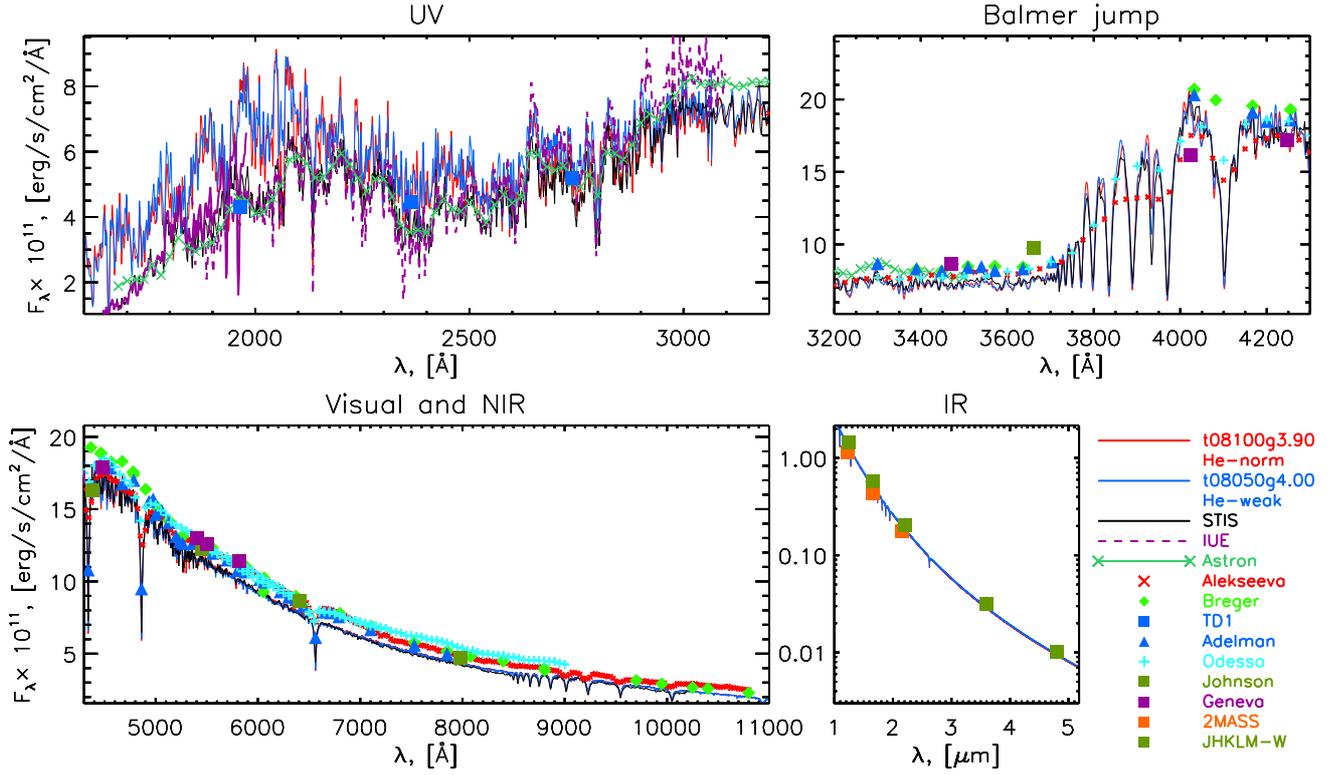}}
\caption{Observed and predicted energy distributions of \bcrb\ for two best-fitted models with
solar and depleted helium contents.}
\label{fig:bcrb-sed}
\end{figure*}

\begin{figure*}
\centerline{\includegraphics[width=\hsize]{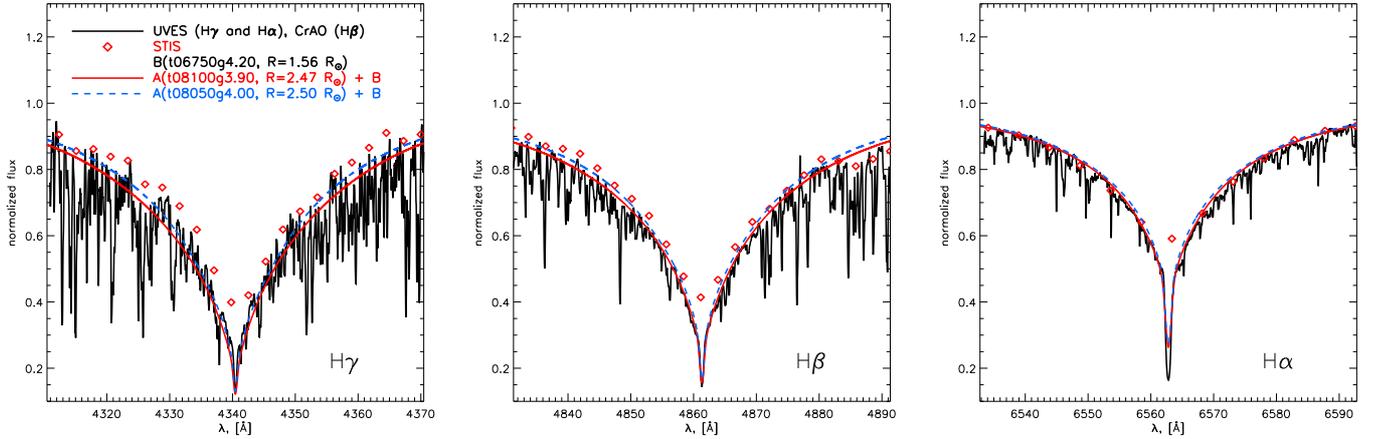}}
\caption{Observed and predicted profiles of hydrogen lines of \bcrb. Full line~--~He norm model; dashed line~--~He-weak model.}
\label{fig:bcrb-h}
\end{figure*}

Available spectra of \bcrb\ made it possible to analyze the stratification of five chemical elements: 
Mg, Si, Ca, Cr, and Fe, using from 8 to 23 neutral and singly-ionized lines per element. We did not perform NLTE
Pr-Nd stratification analysis because of a rather strong magnetic field and an absence of any significant Pr-Nd anomaly 
\citep{2004A&A...423..705R}, which is a manifestation of the rare-earth stratification.
Starting from the $\teff=8000$~K, $\logg=4.3$
homogeneous abundance model, and abundances presented in \citet{2004A&A...423..705R},
the atmospheric parameters were iterated to match 
the observed SED and profiles of hydrogen lines. 
All model atmospheres were computed with $\bs=5.4$~kG surface magnetic field \citep{2008A&A...480..811R}.
The best fitted parameter of the primary
fall in the range $8000<\teff(A)<8050$~K, $3.9<\logg(A)<4.0$ and the respective fits to the SED and hydrogen lines are shown 
on Figs.~\ref{fig:bcrb-sed} and \ref{fig:bcrb-h}. Generally, the $\teff$ of the star
is constrained by the slope of the Paschen continuum and hydrogen lines, while the $\logg$ is best controlled 
by the amplitude of the Balmer jump. 
We find that the contribution of the \bcrb-B is crucial for the Paschen continuum and IR region, and when ignoring
it we were not able to obtain a consistent fit to SED and H-lines with any set of atmospheric parameters.

From Fig.~\ref{fig:bcrb-sed} it is clear that there is a systematic difference between space and ground-based observations.
The STIS, IUE, TD1, and \astron\ data agree very well in the UV region below $\lambda2900$. At the same time, STIS provides
less flux in the visual and NIR than all ground-based photometric and spectrophotometric observations, which in turn agree fairly well
with each other, with the exception of NIR (where some deviation can be seen, especially in the case of data from the
Odessa catalog) and $\ll3800-5000$ region. 
Most likely this is due to the different calibration methods applied to the observed data, and it is hard to give
a preference to any of these datasets.
Because STIS provides most homogeneous sets of observations covering a wide wavelength range from UV to NIR, 
we used it as a reference in the model-fitting process.

Two problems still remain. First, none of the models can reasonably fit the UV spectrum of the star below $\lambda2500$.
The secondary component is too faint to influence the combined flux in this spectral region. 
Because all UV data agree well with each other, we are inclined to
conclude that this discrepancy between theoretical and observed fluxes is due to missing opacity in our models.
The second problem arises from the fit to H-lines. Only $\halpha$ and $\hbeta$ are fit satisfactory.
The computed profiles of $\hgamma$ are always wider than the observed ones even for
He-weak model with lower $\teff=8050$~K that demonstrate somewhat better agreement for these two particular 
lines and worse for $\halpha$. 
By increasing $\teff$ of the secondary within the limits given by \citet{2010A&A...512A..55B}, one can 
obtain a slightly lower $\teff$ for the primary and thus narrower profiles of $\hgamma$, respectively, 
but the fit to both NIR fluxes and particularly $\halpha$ profile then becomes noticeably worse.
By integrating STIS fluxes and theoretical fluxes of the $A=[8050,4.0,2.50\pm0.07]$, $B=[6750,4.2,1.56]$ model,
we estimated a corresponding uncertainty in the effective temperature due to
a difference between observed and predicted fluxes to be $\Delta\teff=50$~K. This means that
the true $\teff$ of \bcrb\ is probably slightly overestimated in our analysis. On the other hand, an uncertainty
in He abundance results in a similar $50$~K error, thus accounting for both, these effects
could lower a temperature of the star by $100$~K compared to the temperature derived in this study.

The final adopted parameters are (component$=[\teff (K),\logg (dex),R (\Rsun)]$): $A=[8050,4.0,2.50\pm0.07]$
for the He-weak model and $A=[8100,3.9,2.47\pm0.07$ for the He-normal 
(e.g., with solar helium composition) model. The parameters of the secondary were taken from \citet{2010A&A...512A..55B}:
$B=[6750,4.2,1.56]$. We used the Hipparcos parallax of $\pi=28.60\pm0.69$~mas from \citet{leeuwen}. 
The errors of the derived radii results from the stellar parallax uncertainty.

Using ground-based spectrophotometry led to a noticeable increase in the stellar radii. For instance,
the fit to the data obtained by \citet{1976ApJS...32....7B} requires $A=[8050,4.0,2.70\pm0.07]$
for He-weak and $A=[8050,3.8,2.65\pm0.07]$ for He-norm models, respectively.
The differences between STIS and ground-based observations correspond to the non-negligible difference in the derived
stellar radii of $\Delta R\approx0.2\Rsun$. This reflects certain problems in the applied flux calibration procedures.
But in spite of the vertical offset, the shape of SEDs remains practically the same, therefore
the effective temperatures derived from the two datasets are very similar.

\begin{figure}
\centerline{
\includegraphics[width=\hsize]{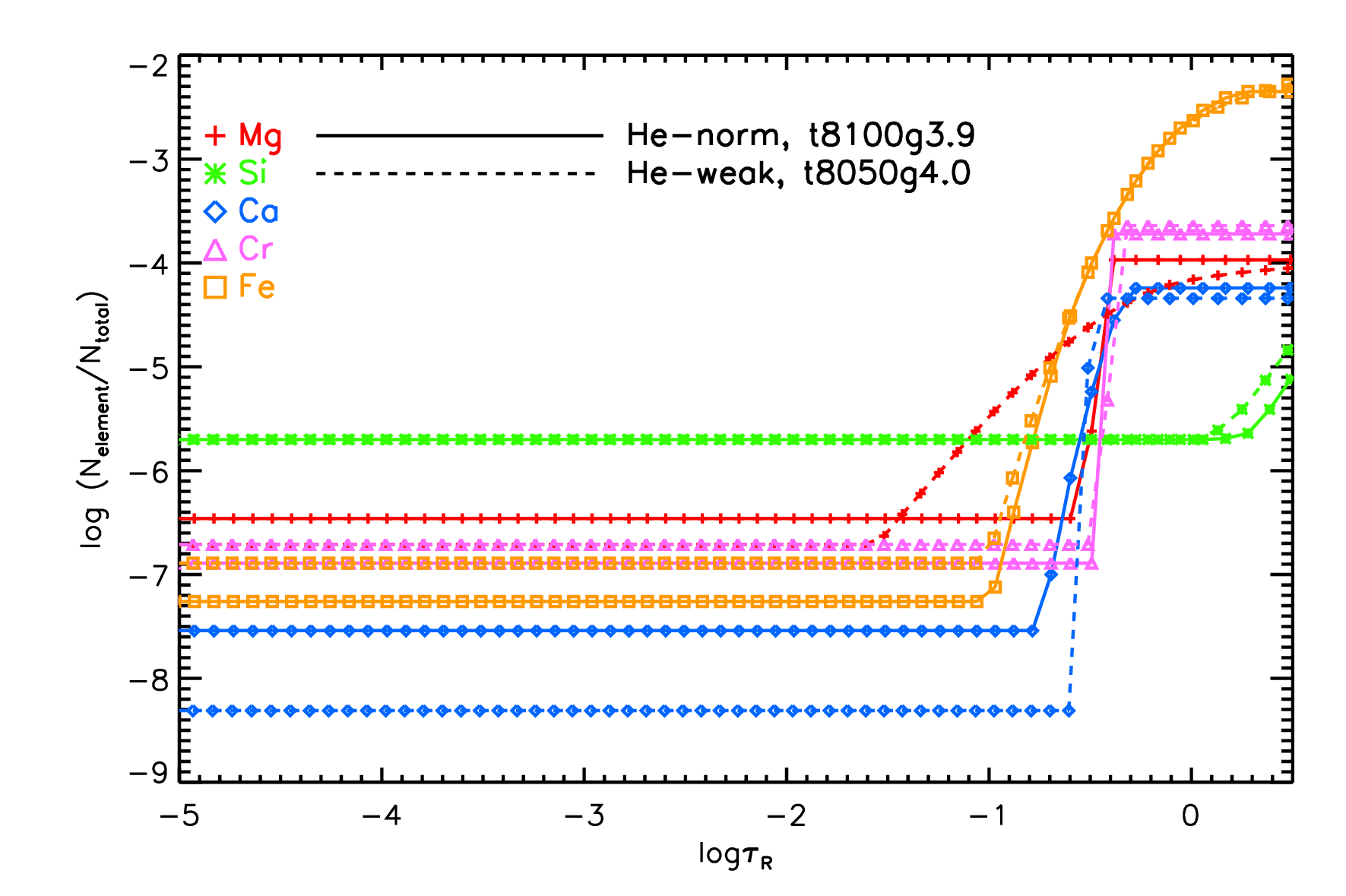}
}
\caption{Depth-dependent distribution profiles of chemical elements in the atmosphere of \bcrb.}
\label{fig:bcrb-str}
\end{figure}

Figure~\ref{fig:bcrb-str} illustrates stratification profiles derived from the final He-norm and He-weak models.
A substantial change in the position of the abundance jump is found for Mg, for which the He-norm model results
in its narrowing and deeper position compared to the He-weak model. Elements Ca and Fe demonstrate
changes in the abundance amplitude in upper atmospheric layers, but we find that solution for Ca is quite
uncertain in the highest layers with estimated errors on the order of $\pm 1$~dex or higher. Stratifications of
Si and Cr stay almost unchanged, with Si accumulating at much deeper photospheric
layers than what was found for other elements. The changes in stratification profiles observed in Fig.~\ref{fig:bcrb-str} 
are a combined effect of decreasing He content and $\teff$ by $50$~K, i.e. not entirely due to He content.
The effect of decreasing He abundance manifests itself in the increased $\logg$ by $0.1$~dex needed to fit the SED of the star.
Thus, accounting for a possible peculiar He content does have a well determined effect on the stellar parameter 
determination, and we suggest to compute models with strongly deficient He content (i.e. in the same way it is done in this research)
when analyzing cool CP stars where true He abundance cannot be derived spectroscopically.

\subsection{\gequ\ (HD~201601)}

Another well known roAp star with a rich spectrophotometric dataset available is \gequ. 
Similar to \bcrb, recent interferometric observations by \citet{2011A&A...526A..89P} provide a
radius of the star $R=2.20\pm0.12\Rsun$. It is also a binary system \citep[see][]{2002A&A...384..180F}
with \gequ-B having much cooler temperature ($\teff(B)=[4570,4833]$~K)
and thus smaller radius than  \bcrb-B, which then translates to the contribution of about $6$\% to the total flux of the system
\citep{2011A&A...526A..89P}. The authors also estimated the effective temperature of the system
$\teff(A+B)=7364\pm235$~K, and thus $\teff(A)=7253$~K. This is much cooler than the temperature of $\teff=7700$~K adopted for \gequ\ in 
the latest detailed spectroscopic study by \citet{2002A&A...384..545R}.

\begin{figure*}
\centerline{\includegraphics[width=\hsize]{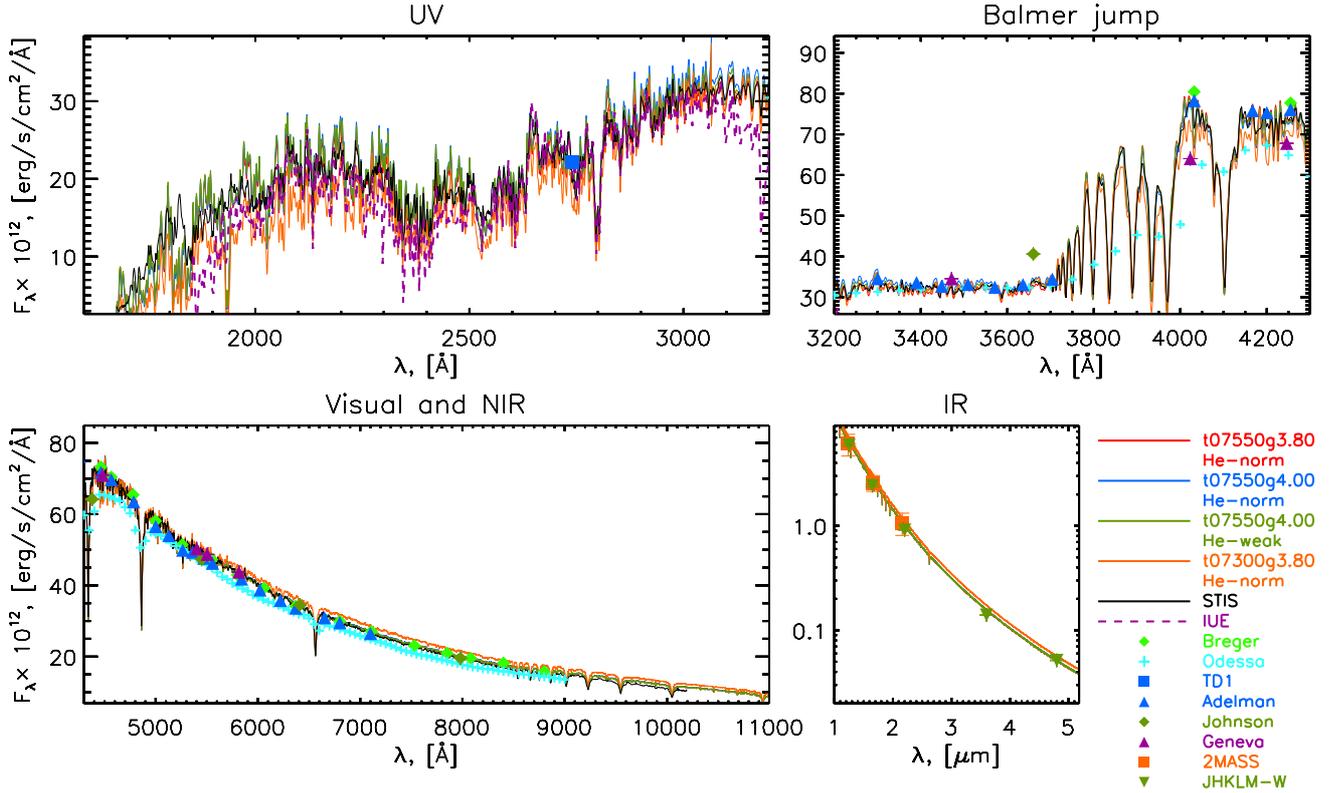}}
\caption{Observed and predicted energy distributions of \gequ\ for two best-fitted models with
solar and depleted helium contents.}
\label{fig:gequ-sed}
\end{figure*}

Due to extremely sharp spectral lines and weaker magnetic field which minimize the blending effects, \gequ\ provides a possibility to
perform a stratification analysis for $18$ chemical elements:
{Na}{}, {Mg}{}, {Si}{}, {Ca}{}, {Sc}{}, {Ti}{}, {V}{}, {Cr}{}, {Mn}{}, {Fe}{},
{Co}{}, {Ni}{}, {Sr}{}, {Y}{}, {Zr}{}, {Ba}{}, {Pr}{}, {Nd}{}.
This represents the most complete empirical study of element diffusion in an atmosphere of a single star to date.
Starting from the initial model with $\teff=7700$~K, $\logg=4.2$ from \citet{1997A&A...322..234R}, we iterated 
stratification analysis
and model atmosphere calculations until a minimal deviation between observed spectrophotometric observations and model
fluxes was obtained. The surface magnetic field introduced in model atmosphere computations was $\bs=4$~kG.
In {Sc}{}, {V}{}, {Mn}{}, {Co}{} stratification calculations laboratory data for hyperfine structure ({\it hfs}) were taken into account.
They were extracted from the following papers: \citet{1992PhyS...46...45V} (\ion{Sc}{II}), \citet{2011PhyS...84e5301A} (\ion{V}{II}), 
\citet{2005ApJS..157..402B} (\ion{Mn}{I}), \citet{1999MNRAS.306..107H} (\ion{Mn}{II}), \citet{1996ApJS..107..811P} (\ion{Co}{I}), \citet{2010MNRAS..401.1334B}
(\ion{Co}{II}). Hyperfine structure for \ion{Pr}{ii} was also included in the {Pr}{} stratification analysis \citep[see][]{2009A&A...495..297M} with the 
{\it hfs}-constants from \citet{ginibre}. 
In the case of He-norm setting, we find that models with $A=[7600,4.0,2.04\pm0.05]$, 
$B=[4500-4750,4.5,0.65-0.70]$ provide the best fit to the observed SED applying
parallax $\pi=27.55\pm0.62$ mas from \citet{leeuwen}.
Assuming $B=[5000,4.5,0.80]$ results in a higher temperature of the primary
$A=[7700,4.0,1.95\pm0.05]$. To fit the Balmer lines, however, a lower effective temperatures of about $\teff=7300$~K is needed.
Therefore, to maintain a satisfactory fit to both Balmer lines and the SED, a slightly cooler $A=[7550,4.0,2.07\pm0.05]$ was chosen.
We note that formally the model with $\logg=3.8$ provides a better fit to SED ($\chi^2=1.19$) than the model with $\logg=4.0$ ($\chi^2=1.40$)
with a difference visible at the Balmer jump region (top right plot of Fig.~\ref{fig:gequ-sed}).
But the $\logg=3.8$ then results in too low a mass of the star, $M=0.98\Msun$, which is unreasonable for a main sequence object with $\teff=7550$~K.
No distinction could be made between these two values of $\logg$ from hydrogen line profiles because they are insensitive to the
atmospheric gravity in this temperature range.

From the SED comparison presented in Fig.~\ref{fig:gequ-sed}, it is seen that good agreement 
between observations and model predictions is obtained for all spectral intervals represented  from the UV to IR.
Ground-based spectrophotometric data are also found to be consistent with the STIS observations,
except observations obtained by the Odessa research group which are systematically lower.
The derived effective temperatures from the SED are reasonably close to those of \citet{1997A&A...322..234R}. 
This is because in the latter work the authors fitted observations of \citet{1989A&AS...81..221A}, which required
$\teff=7700$~K to match the slope of the Paschen continuum. 

\begin{figure*}
\centerline{\includegraphics[width=\hsize]{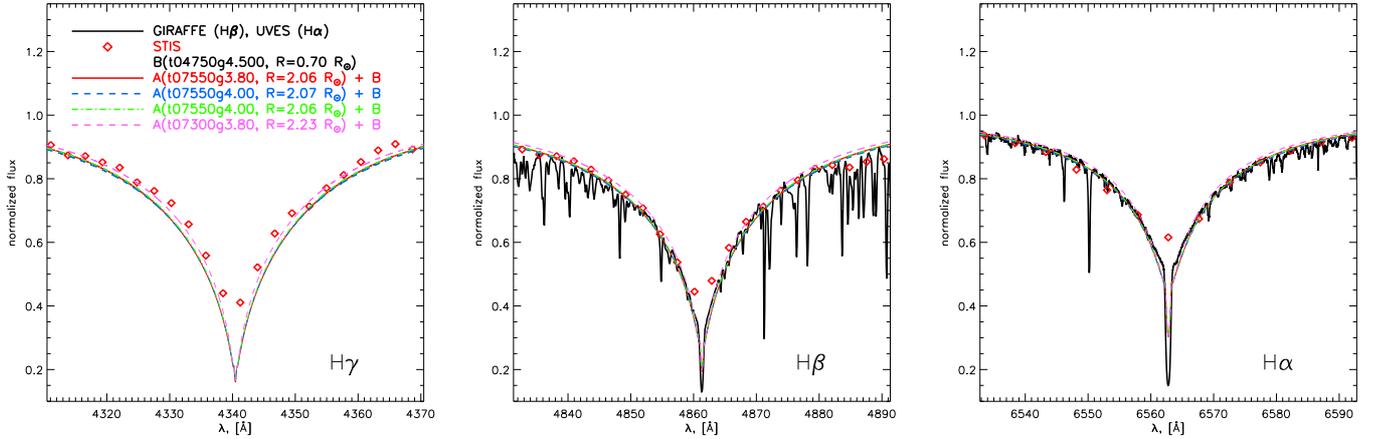}}
\caption{Observed and predicted profiles of hydrogen lines of \gequ. Dash-dotted lines correspond to He-weak model.}
\label{fig:gequ-h}
\end{figure*}

Figure~\ref{fig:gequ-h} compares the observed and predicted profiles of hydrogen lines.
The normalization of 
UVES data at the $\hbeta$ line was found to suffer considerably from the data processing inaccuracies. Therefore we
used observed profile taken with GIRAFFE instrument at SAAO. The normalization
of the STIS data in the same region is also not trivial but easier to perform due to its lower resolution
and a wide wavelength coverage. On the other hand, $\halpha$ line of STIS data is still satisfactorly fitted
with $\teff=7550$~K models, though the UVES profile lies systematically higher, and thus lower $\teff$
is needed to fit it. The underabundance of He formally resulted in an increase in $\logg$ from $3.8$ to $4.0$
so that the He-weak model parameters are $A=[7550,4.0,2.06\pm0.05]$.

As follows from the results presented above, the most noticeably affected parameter is $\logg$, which
underwent changes from $4.2$ to $3.8-4.0$ depending upon the helium abundance used. A value of
$\logg=4.2$ was used in \citet{1997A&A...322..234R,2002A&A...384..545R} to fit atomic lines using atmospheric models
with a scaled solar abundances, while in our analysis we used model atmospheres and fluxes
computed with realistic abundances of the star. Finally, the use of STIS data allowed us to
achieve a more detailed fit to the region of the Balmer jump, whose amplitude is very sensitive to the
value of $\logg$.

\begin{figure}
\centerline{\includegraphics[width=\hsize]{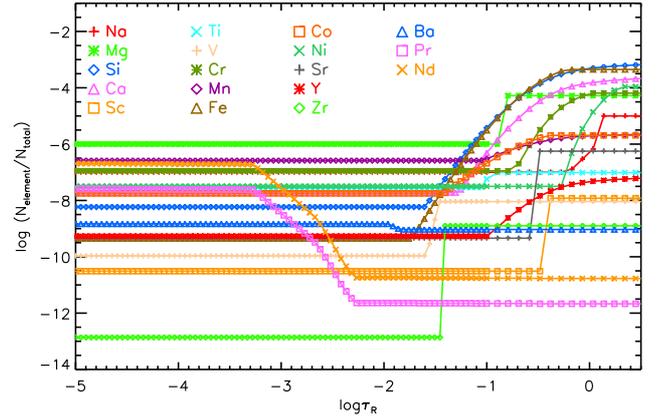}}
\caption{Depth-dependent distribution profiles of chemical elements in the atmosphere of \gequ.}
\label{fig:gequ-str}
\end{figure}

The vertical distribution profiles of all $18$ chemical elements derived with $\teff=7550$~K, $\logg=3.8$, He-norm model
are presented in Fig.~\ref{fig:gequ-str}.
All but REE elements (i.e. {Pr}{} and {Nd}{}) are found to accumulate at layers just above the photosphere.
For elements {Ti} and {Ba}, stratification is not well constrained and probably is
not present in the atmosphere of this star at all. A cloud of {Pr}{} and {Nd}{} is observed at high
atmospheres and the abundance gradients of these elements, together with {Fe}{} and {Si}{},
are the largest among the studied elements.

\subsection{\lib\ (HD~137949)}

In the present investigation \lib\ is the last target analyzed. It is a bright roAp star with many references available regardless
its spectroscopic and photometric properties. Unfortunately, no high-resolution spectroscopic energy distribution is available
similar to what
is provided by STIS, therefore a fit was done to the ground-based spectrophotometric data from \citet{1989A&AS...81..221A}. 
Our fitting began with $\teff=7550$~K, $\logg=4.3$ model used by \citet{2004A&A...423..705R,2008A&A...480..811R}. 
Stratification of {Si}{}, {Fe}{}, {Ca}{}, and {Cr}{} was derived and accounted for in the model atmosphere calculations. 
Average abundances of the REE are higher in \lib\ than in \gequ\ by more than an order of magnitude. The stronger magnetic 
field of $\bs=5$~kG \citep{2008A&A...480..811R}, together with high REE abundances, results in extremely strong 
blending that prevents the proper choice of spectral lines 
for any stratification study of other elements. It is also the reason we could not perform NLTE stratification analysis 
of Pr and Nd based on
equivalent widths and pseudomicroturbulence approximation of magnetic effects. It could only be done by properly including the
magnetic field in NLTE line formation code. Pr-Nd stratification is expected to be rather strong in \lib\ because of the highest
Pr-Nd anomaly observed in this star \citep{2004A&A...423..705R}; however, at present step we limit ourselves by taking
the average observed REE anomalies into account.

\begin{figure*}
\centerline{\includegraphics[width=\hsize]{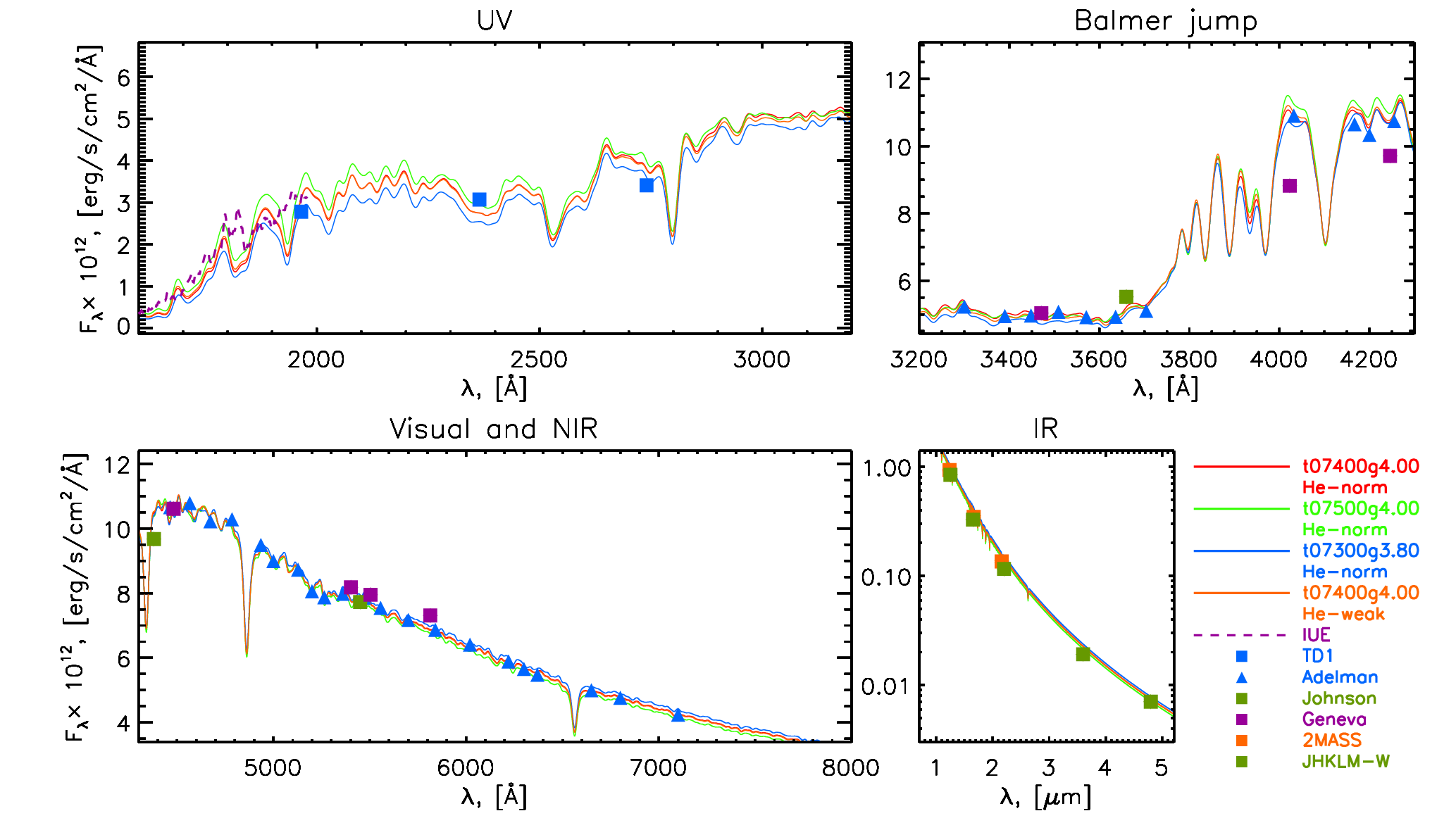}}
\caption{Observed and predicted energy distributions of \lib\ for two best-fitted models with
solar and depleted helium contents.}
\label{fig:lib-sed}
\end{figure*}

The best-fit parameters are $[7400,4.0,2.13\pm0.13]$ for both He-norm and He-weak models. As shown in Fig.~\ref{fig:lib-sed},
also $[7300,3.8,2.19\pm0.13]$ and $[7500,4.0,2.06\pm0.13]$ models provide a reasonable fit, but with more
pronounced deviations between
the observed and predicted SEDs at short and long wavelengths.

\begin{figure*}
\centerline{\includegraphics[width=\hsize]{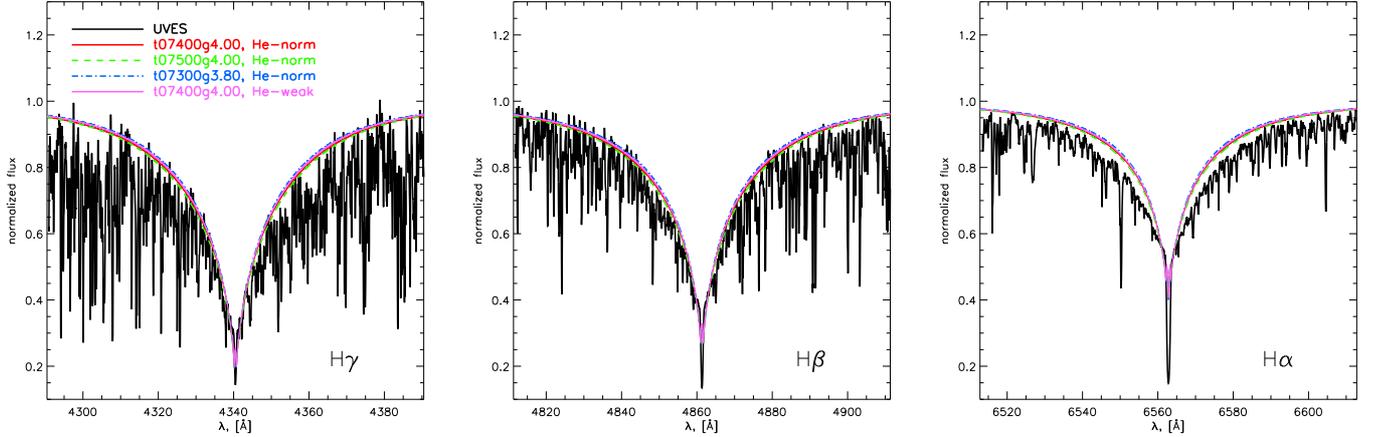}}
\caption{Observed and predicted profiles of hydrogen lines of \lib.}
\label{fig:lib-h}
\end{figure*}

Similar to other cases described above, for \lib\ we also found it difficult to fit the Balmer $\halpha$ line with 
the model parameters required
to fit the SED (see Fig.~\ref{fig:lib-h}). A higher $\teff$ is required to fit the observed profile of this line. On the other
hand, $\hbeta$ and $\hgamma$ lines are a satisfactory fit with the $\teff=7400$~K model.

\begin{figure*}
\centerline{
\includegraphics[width=0.45\hsize]{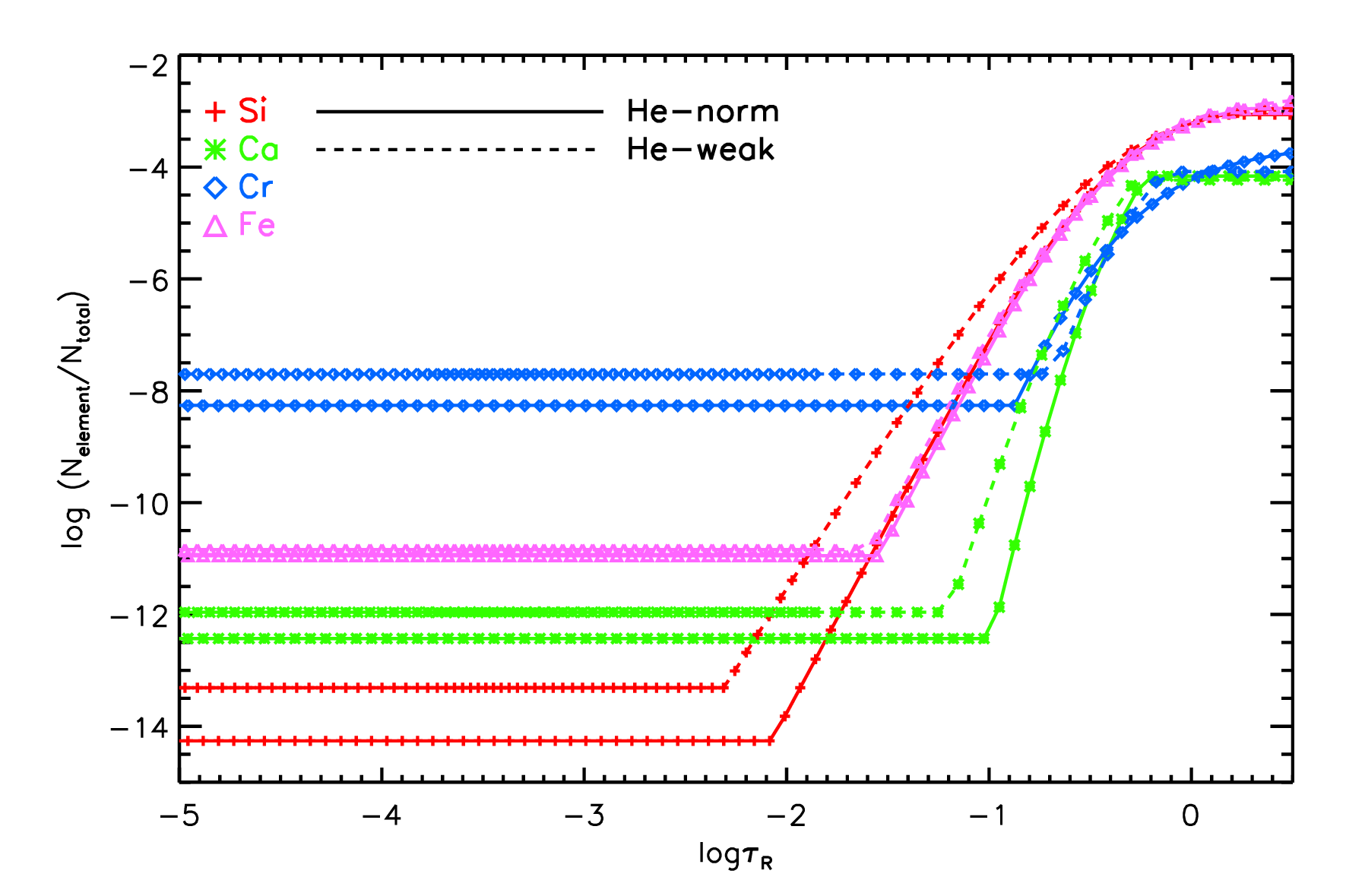}
\includegraphics[width=0.45\hsize]{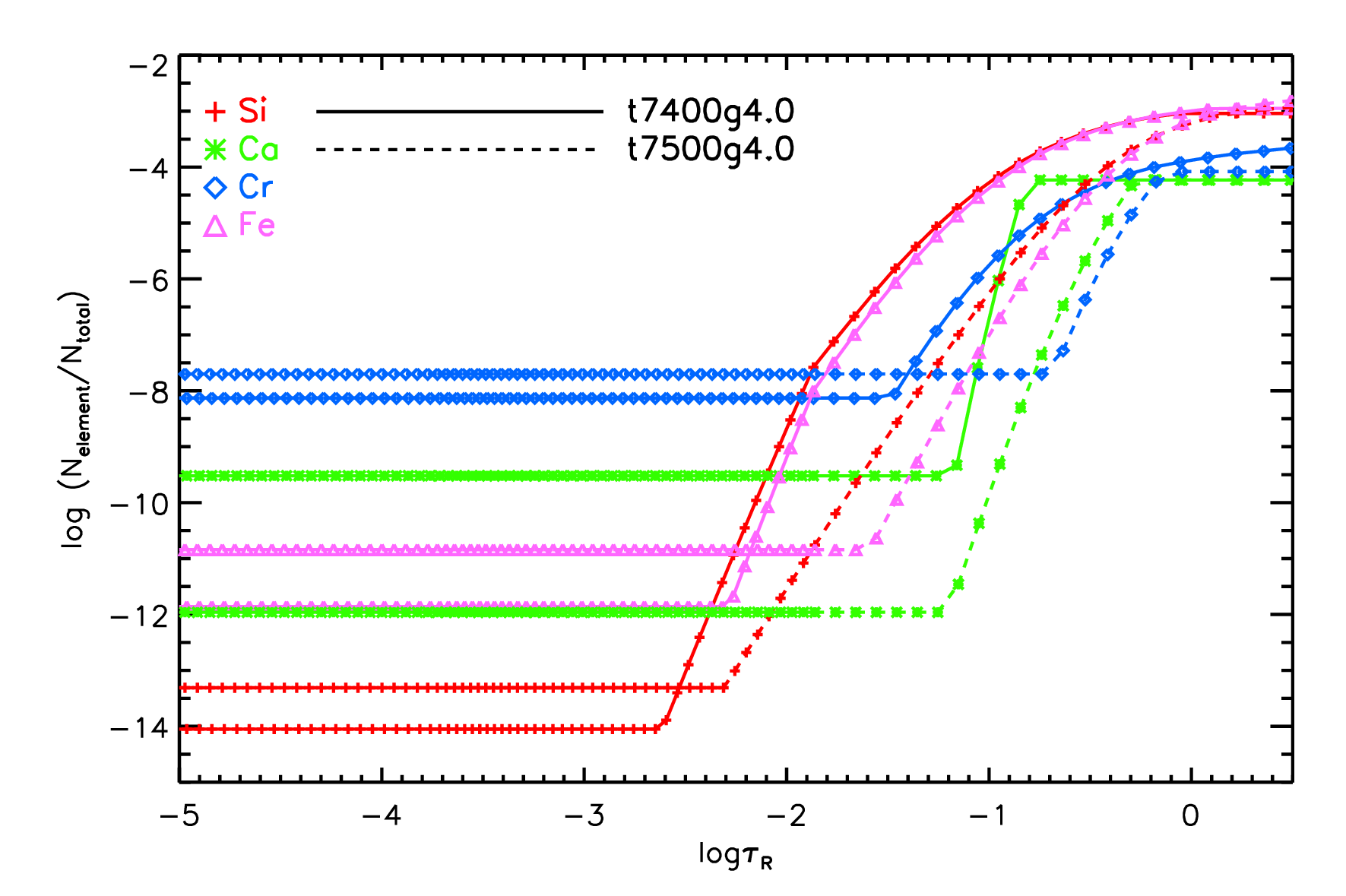}
}
\caption{Depth-dependent distribution profiles of chemical elements in the atmosphere of \lib\ for He-norm and He-weak
$\teff=7400$~K, $\logg=4.0$ models (left panel), and two He-weak $\teff=7400$~K, $\logg=4.0$, and $\teff=7500$~K, $\logg=4.0$
models (right panel).}
\label{fig:lib-str}
\end{figure*}

The depth-dependent distribution of elements derived using the $[7400,4.0]$ He-norm and He-weak models is 
shown on Fig.~\ref{fig:lib-str}. As expected, in both cases the stratification profiles are very similar, with
large abundance jumps at deep atmospheric layers.

\section{Discussion and conclusions}
\label{sec:discussion}

\begin{table*}
\caption{Fudamental parameters of Ap stars.}
\label{tab:results}
\begin{scriptsize}
\begin{center}
\begin{tabular}{l|ccccc|ccccc|cc}
\hline\hline
\multirow{2}{*}{Name}         & \multicolumn{5}{|c|}{He-norm} & \multicolumn{5}{|c|}{He-weak} & \multirow{2}{*}{$R_{\rm inter}^{\mathrm{a}}$, $\Rsun$} & \multirow{2}{*}{$\bs$, kG}\\
             & $\teff$, K & $\logg$ & $R$, $\Rsun$ & $L$, $\Lsun$ & $M$, $\Msun$ & $\teff$, K & $\logg$ & $R$, $\Rsun$ & $L$, $\Lsun$ & $M$, $\Msun$ &&\\
\hline
HD~137909    & $8100$ & $3.9$ & $2.47\pm0.07$ & $23.69\pm1.93$ & $1.77\pm0.51$ & $8050$ & $4.0$ & $2.50\pm0.07$ & $23.67\pm1.91$ & $2.28\pm0.65$ & $2.63\pm0.09^{1}$ & $5.4$\\ 
\hline
HD~201601    & $7550$ & $4.0$ & $2.07\pm0.05$ & $12.56\pm0.94$ & $1.56\pm0.44$ & $7550$ & $4.0$ & $2.06\pm0.05$ & $12.44\pm0.93$ & $1.55\pm0.43$ & $2.20\pm0.12^{2}$ & $4.0$\\ 
\hline
HD~137949              & $7400$ & $4.0$ & $2.13\pm0.13$ & $12.27\pm1.83$ & $1.66\pm0.58$ & $7400$ & $4.0$ & $2.13\pm0.13$ & $12.27\pm1.83$ & $1.66\pm0.58$ &                 & $5.0$\\ 
\hline
HD~24712$^{\rm 3}$     & $7250$ & $4.1$ & $1.77\pm0.04$ & $7.81\pm0.57$  & $1.44\pm0.40$ &        &       &               &                &               &                 & $3.1$\\
\hline
HD~101065$^{\rm 4}$    & $6400$ & $4.2$ & $1.98\pm0.03$ & $5.92\pm0.37$  & $2.27\pm0.59$ &        &       &               &                &               &                 & $2.3$\\
\hline
\multirow{2}{*}{HD~103498$^{\rm 5}$}    & $9300$ & $3.5$ & $4.56\pm0.77$ & $140.28\pm50.39$  & $2.40\pm1.36$ &        &       &               &                &               &                 &\\
                       & $9500$ & $3.6$ & $4.39\pm0.75$ & $141.57\pm51.35$  & $2.80\pm1.60$ &        &       &               &                &               &                 &\\
\hline
HD~128898$^{\rm 6}$    & $7500$ & $4.1$ & $1.94\pm0.004$ & $10.74\pm0.33$  & $1.73\pm0.41$ &        &       &               &                &               &  $1.967\pm0.066^{7}$           & $\sim2.0$\\
\hline
\end{tabular}
\end{center}
\end{scriptsize}
The error bars of $M$ and $R$ are average errors computed taking parallax uncertainity and additionally
assuming errors of $\Delta\teff=50$~K and $\Delta\logg=0.1$~dex into account. Larger errors found for HD~103498 result from a
large parallax uncertainity.\\
$^{\mathrm{a}}$ Radius derived by means of interferometry.\\
$^1$ \citet{2010A&A...512A..55B}; $^2$ \citet{2011A&A...526A..89P}; $^3$ \citet{2009A&A...499..879S}; $^4$ \citet{2010A&A...520A..88S};
$^5$ \citet{2011MNRAS.417..444P}; $^6$ \citet{2009A&A...499..851K};\\
$^7$ \citet{2008MNRAS.386.2039B}
\end{table*}

In an attempt to construct self-consistent models of Ap-star atmospheres, we performed a simultaneous model fit to a number
of observed datasets, such as photometric wide-band fluxes, low- and middle-resolution spectrophotometric observations,
and high-resolution spectra. Ideally, all of them must be fitted with a single model atmosphere and abundance
pattern. The results of the present investigation suggest that this is indeed possible to accomplish by applying a realistic
chemical composition and appropriate model atmospheres that account for the peculiar nature of CP stars.

However, a few problems still exist. The most disturbing one is a systematic discrepancy between effective temperatures
derived separately from SED and H-lines. In the case of \bcrb, only $\halpha$ and $\hbeta$ agree with SED, but $\hgamma$
requires lower $\teff$. For \lib\ the situation is the opposite with $\hbeta$ and $\hgamma$ in agreement with
the $\teff$ derived from SED, but $\halpha$ clearly requires higher $\teff$.
A good match to SED, $\halpha$, and $\hbeta$ was obtained for \gequ, though no high-resolution data was available
for the $\hgamma$ line region.
Since at least two out of three hydrogen lines always
agree with the atmospheric parameters derived from fitting the SED, it is likely that we face the usual problem of
continuum normalization of the hydrogen lines in cross-dispersed echelle observations. 
Thus, one should always be cautious when determining
atmospheric parameters solely based on the fit to hydrogen lines.

The abundance and stratification analysis was done using carefully selected spectral lines with accurately
known transition parameters, and yet there seems to be a missing absorption in the models of \bcrb\ 
in UV region blueward $\lambda2500$ (see top left panel of Fig.~\ref{fig:bcrb-sed}). A secondary component is by far too 
faint and cool to influence the flux in that spectral region.

For \bcrb\ we find a noticeable deviation between the space- and ground-based observations of the SEDs
in the visual and NIR regions. A natural reason could be a discrepancy between different flux calibration schemes, which by default
are more challenging in the case of ground-based observations. This deviation is seen as a vertical offset in the observed fluxes.
It does not affect the determination of $\teff$ (Fig.~\ref{fig:bcrb-sed}), but does affect the radius estimate by about $0.2\Rsun$.
This is quite a noticeable uncertainty that highlights once again the need to have accurately calibrated fluxes for stellar
parameter determinations.

All previous spectroscopic studies resulted in the $\logg$ values that were substantially over-estimated, as shown by
direct fits presented on Figs.~\ref{fig:bcrb-sed},~\ref{fig:gequ-sed}, and~\ref{fig:lib-sed}. Our values come
mainly from the detailed fit to the amplitude and slope of the Balmer jump based on realistic model atmospheres.
Previously, only H-lines, photometric colors and/or low-resolution spectrophotometry by \citet{1989A&AS...81..221A}
were used in combination with scaled-solar metallicity models that did not account for individual abundances, magnetic fields, and
improved REE opacity.

We find that possible He depletion of Ap-star atmospheres has minor effect on determining  element distributions and $\logg$,
at least for the hottest star considered here, \bcrb (see Fig.~\ref{fig:bcrb-str}).
Formally, an underabundance of He results in choosing a slightly higher $\logg$ by about $0.1$~dex and a cooler
$\teff$ by $\approx50$~K. This change is small and well within the error bars of our fitting procedure. The choice of temperature 
has a somewhat stronger or at least a comparable impact on the derived stratification profiles of chemical elements, 
as can be seen from the example of \lib\ illustrated in left- and right-hand panels of Fig.~\ref{fig:lib-str}.
In particular, a change of $\Delta\teff=100$~K influences both positions and amplitudes of the abundance jumps.
It is important to realize, however, that even if we are not absolutely precise in deriving $\teff$ and know nothing
about the real He content of the Ap-star atmospheres, the inferred element stratifications are still informative and stable.
Additional uncertainties could result from particular limitations in our fitting procedure 
(e.g., step-like approximation of stratification profiles, a small number of spectral lines used, etc.), and thus
the presented analysis can only provide a general picture of the element distributions in the atmospheres of investigated stars. 
Still, this is enough to reveal an accumulation of a particular element in the lower or upper atmospheric layers, 
determine the amplitudes of abundance jumps, and estimate
uncertainties in the derived distributions due to assumed model parameters~--~questions we were mainly interested in.

Interferometry has recently brought an entirely new possibility of measuring the radii of main-sequence stars on a model-independent
and regular basis, though its modern instrumental capabilities are limited to bright nearby stars.
Our model-based comparison of the
observed and predicted SEDs led to good agreement (within error bars) with interferometric estimates
as summarized in Table~\ref{tab:results}. The table also includes calculated masses and luminosities from derived
effective temperatures, radii, and gravities. The corresponding error bars were calculated
using individual uncertainties of stellar parallaxes along with typical uncertainties of $\Delta\teff=50$~K and
$\Delta\logg=0.1$~dex associated with the fit of SEDs and hydrogen line profiles. Large error bars found for
HD~103498 are entirely due to a large parallax uncertainty ($\pi=3.37\pm0.56$ mas given by \citet{leeuwen}).

\begin{figure}
\centerline{
\includegraphics[width=8cm]{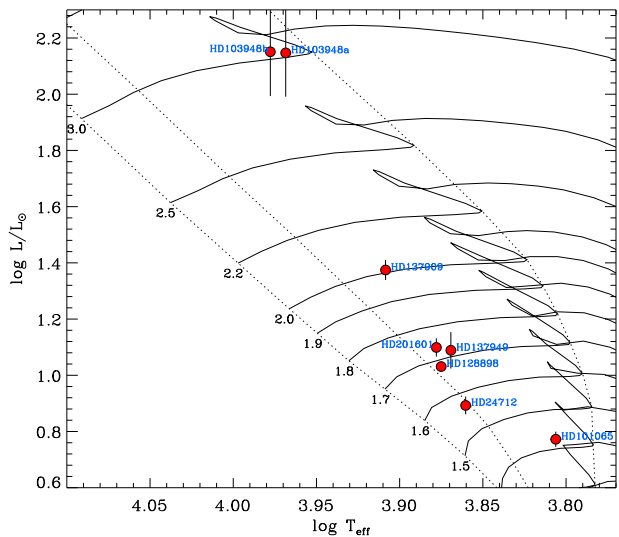}
}
\caption{Predicted location of Ap stars in the H-R diagram for the luminosity and temperature estimates from 
Table~\ref{tab:results}. Theoretical evolutionary tracks were adopted from \citet{2000A&AS..141..371G} 
for the bulk metallically of $Z=0.018$. The dotted lines show ZAMS, $50$\% fractional main sequence age line, and TAMS.}
\label{fig:hr}
\end{figure}

Figure~\ref{fig:hr} shows the positions of stars from Table~\ref{tab:results} in the H-R diagram along with
theoretical evolutionary tracks. This comparison allows deriving stellar masses and ages under a reasonable assumption 
about the bulk stellar metallicity. Our spectroscopically derived masses generally show good agreement
with the theoretical values predicted by evolutionary models within estimated error bars. The only exception
is HD~101065 for which evolutionary models predict $M=1.48$~$\Msun$, while \citet{2010A&A...520A..88S}
estimates $2.27\pm0.59$~$\Msun$. In this particular case, however, the derived $\logg$ was very sensitive
to the assumed line opacity of REE elements that were found to affect model fluxes in the Balmer jump
region in a dramatic way. Because of NLTE effects and the possible incompleteness of REE line lists,
the published value of $\logg=4.2$ derived from the fit to the observed SED
could easily be overestimated, and a somewhat lower $\logg=4.0$ would give a consistent mass estimate.

\begin{acknowledgements}
We thank Dr. Igor Savanov for kindly providing us with $\hbeta$ spectrum of \bcrb.

This work was supported by the following grants: Deutsche Forschungsgemeinschaft (DFG)
Research Grant RE1664/7-1 to DS and Basic Research Program of the Russian Academy of Sciences ``Nonstationary phenomena in the Universe''
to TR. OK is a Royal Swedish Academy of Sciences Research Fellow supported by grants from the Knut and Alice Wallenberg Foundation and from the Swedish Research Council.

Some of the data presented in this paper were obtained from the Mikulski Archive for Space Telescopes (MAST). 
STScI is operated by the Association of Universities for Research in Astronomy, Inc., under NASA contract NAS5-26555. 
Support for MAST for non-HST data is provided by the NASA Office of Space Science via grant NNX09AF08G 
and by other grants and contracts.

We also acknowledge the use of electronic databases (VALD, SIMBAD, NASA's ADS) and 
cluster facilities at the Vienna Institute for Astronomy and Georg August University G\"ottingen.
\end{acknowledgements}



\begin{thebibliography}{}
\bibitem[Adelman et al.(1989)]{1989A&AS...81..221A} Adelman, S.~J., Pyper, D.~M., Shore, S.~N., White, R.~E., \& Warren, W.~H., Jr.\ 1989, \aaps, 81, 221
\bibitem[Alekseeva et al.(1996)]{1996BaltA...5..603A} Alekseeva, G.~A., Arkharov, A.~A., Galkin, V.~D., et al.\ 1996, Baltic Astronomy, 5, 603
\bibitem[Armstrong et al.(2011)]{2011PhyS...84e5301A} Armstrong, N.~M.~R., Rosner, S.~D., \& Holt, R.~A.\ 2011, Phys. Scripta, 84, 055301  
\bibitem[Bergemann et al.(2010)]{2010MNRAS..401.1334B} Bergemann, M., Pickering, J.~C., \& Gehren, T. \ 2010, \mnras, 401, 1334  
\bibitem[Blackwell-Whitehead et al.(2005)]{2005ApJS..157..402B} Blackwell-Whitehead, R.~J., Pickering, J.~C., Pearse, O., \& Nave, J.\ 2003, \apjs, 157, 402  
\bibitem[Boiarchuk(1986)]{1986INTSA..31..198B} Boiarchuk, A.~A.\ 1986, Itogi Nauki i Tekhniki Seriia Astronomiia, 31, 198 
\bibitem[Boksenberg et al.(1973)]{1973MNRAS.163..291B} Boksenberg, A., Evans, R.~G., Fowler, R.~G., et al.\ 1973, \mnras, 163, 291
\bibitem[Breger(1976)]{1976ApJS...32....7B} Breger, M.\ 1976, \apjs, 32, 7
\bibitem[Bruntt et al.(2010)]{2010A&A...512A..55B} Bruntt, H., Kervella, P., M{\'e}rand, A., et al.\ 2010, \aap, 512, A55
\bibitem[Bruntt et al.(2008)]{2008MNRAS.386.2039B} Bruntt, H., North, J.~R., Cunha, M., et al.\ 2008, \mnras, 386, 2039
\bibitem[Cowley et al.(2007)]{2007MNRAS.377.1579C} Cowley, C.~R., Hubrig, S., Castelli, F., González, J.~F., \& Wolff, B. \ 2007, \mnras, 377, 1579
\bibitem[Dekker et al.(2000)]{DOK00} Dekker, H., D'Odorico, S., Kaufer, A., Delabre, B., \& Kotzlowski, H. 2000, proc. SPIE, 4008, 534
\bibitem[Fabricius et al.(2002)]{2002A&A...384..180F} Fabricius, C., H{\o}g, E., Makarov, V.~V., et al.\ 2002, \aap, 384, 180
\bibitem[Ginibre(1989)]{ginibre} Ginibre, A. \ 1989, Phys. Scripta, 39, 710 
\bibitem[Girardi et al.(2000)]{2000A&AS..141..371G} Girardi, L., Bressan, A., Bertelli, G., \& Chiosi, C.\ 2000, \aaps, 141, 371
\bibitem[Holt et al.(1999)]{1999MNRAS.306..107H} Holt, R.~A., Scholl, T.~J., \& Rosner, S.~D. \ 1999, \mnras, 306, 107  
\bibitem[Khan \& Shulyak(2006a)]{2006A&A...448.1153K} Khan, S.~A., \& Shulyak, D.~V.\ 2006a, \aap, 448, 1153
\bibitem[Khan \& Shulyak(2006b)]{2006A&A...454..933K} Khan, S.~A., \& Shulyak, D.~V.\ 2006b, \aap, 454, 933
\bibitem[Kochukhov et al.(2009)]{2009A&A...499..851K} Kochukhov, O., Shulyak, D., \& Ryabchikova, T.\ 2009, \aap, 499, 85
\bibitem[Kochukhov(2007)]{synthmag07}Kochukhov, O. 2007, in \textit{Physics of Magnetic Stars}, eds.D.O.~Kudryavtsev and I.I~Romanyuk, Nizhnij Arkhyz., p.109
\bibitem[Komarov et al.(1995)]{1995OAP.....8....3K} Komarov, N.~S., Dragunova, A.~V., Belik, S.~I., et al.\ 1995, Odessa Astronomical Publications, 8, 3 
\bibitem[Krti{\v c}ka et al.(2012)]{2012A&A...537A..14K} Krti{\v c}ka, J., Mikul{\'a}{\v s}ek, Z., L{\"u}ftinger, T., et al.\ 2012, \aap, 537, A14
\bibitem[Kupka et al.(1999)]{vald2}Kupka, F., Piskunov, N., Ryabchikova, T. A., Stempels, H. C., \& Weiss, W. W. 1999, \aaps, 138, 119
\bibitem[Kurtz et al.(2006)]{KEM06} Kurtz D. W., Elkin V. G., \& Mathys G. \ 2006, \mnras, 370, 1274
\bibitem[Kurucz(1992)]{1992IAUS..149..225K} Kurucz, R.~L.\ 1992, The Stellar Populations of Galaxies, 149, 225 
\bibitem[Kurucz(1993)]{1993KurCD..13.....K} Kurucz, R.\ 1993, ATLAS9 Stellar Atmosphere Programs and 2 km/s grid.~Kurucz CD-ROM No.~13.~ Cambridge, Mass.: Smithsonian Astrophysical Observatory, 1993., 13
\bibitem[Leblanc \& Monin(2004)]{2004IAUS..224..193L} Leblanc, F., \& Monin, D.\ 2004, The A-Star Puzzle, 224, 193
\bibitem[Mashonkina et al.(2009)]{2009A&A...495..297M} Mashonkina, L., Ryabchikova, T., Ryabtsev, A., \& Kildiyarova, R.\ 2009, \aap, 495, 297
\bibitem[Mashonkina et al.(2005)]{2005A&A...441..309M} Mashonkina, L., Ryabchikova, T., \& Ryabtsev, A.\ 2005, \aap, 441, 309
\bibitem[Michaud et al.(1979)]{1979ApJ...234..206M} Michaud, G., Martel, A., Montmerle, T., et al.\ 1979, \apj, 234, 206
\bibitem[North et al.(1998)]{1998A&AS..130..223N} North, P., Carquillat, J.-M., Ginestet, N., Carrier, F., \& Udry, S.\ 1998, \aaps, 130, 223
\bibitem[Pandey et al.(2011)]{2011MNRAS.417..444P} Pandey, C.~P., Shulyak, D.~V., Ryabchikova, T., \& Kochukhov, O.\ 2011, \mnras, 417, 444 
\bibitem[Perraut et al.(2011)]{2011A&A...526A..89P} Perraut, K., Brand{\~a}o, I., Mourard, D., et al.\ 2011, \aap, 526, A89
\bibitem[Pickering(1996)]{1996ApJS..107..811P} Pickering, J.~C. \ 2006, \apjs, 107, 811  
\bibitem[Piskunov et al.(1995)]{vald1}Piskunov, N. E., Kupka, F., Ryabchikova, T. A., Weiss, W. W., \& Jeffery, C. S. 1995, \aaps, 112, 525
\bibitem[Ryabchikova et al.(2008)]{2008A&A...480..811R} Ryabchikova, T., Kochukhov, O., \& Bagnulo, S.\ 2008, \aap, 480, 811
\bibitem[Ryabchikova et al.(2007)]{2007A&A...473..907R} Ryabchikova, T., Sachkov, M., Kochukhov, O., Lyashko, D. 2007, \aap, 473, 907
\bibitem[Ryabchikova et al.(2005)]{2005A&A...438..973R} Ryabchikova, T., Leone, F., \& Kochukhov, O.\ 2005, \aap, 438, 973
\bibitem[Ryabchikova et al.(2004)]{2004A&A...423..705R} Ryabchikova, T., Nesvacil, N., Weiss, W.~W., Kochukhov, O., \& St{\"u}tz, C.\ 2004, \aap, 423, 705
\bibitem[Ryabchikova et al.(2002)]{2002A&A...384..545R} Ryabchikova, T., Piskunov, N., Kochukhov, O., et al.\ 2002, \aap, 384, 545
\bibitem[Ryabchikova et al.(1997)]{1997A&A...322..234R}	Ryabchikova, T.~A., Adelman, S.~J., Weiss, W.~W., \& Kuschnig, R.\ 1997, \aap, 322, 234
\bibitem[Savanov \& Kochukhov(1998)]{1998AstL...24..516S} Savanov, I.~S., \& Kochukhov, O.~P.\ 1998, Astronomy Letters, 24, 516
\bibitem[Shulyak et al.(2010a)]{2010A&A...524A..66S} Shulyak, D., Krti{\v c}ka, J., Mikul{\'a}{\v s}ek, Z., Kochukhov, O., L{\"u}ftinger, T.\ 2010a, \aap, 524, A66
\bibitem[Shulyak et al.(2010b)]{2010A&A...520A..88S} Shulyak, D., Ryabchikova, T., Kildiyarova, R., \& Kochukhov, O.\ 2010b, \aap, 520, A88
\bibitem[Shulyak et al.(2009)]{2009A&A...499..879S} Shulyak, D., Ryabchikova, T., Mashonkina, L., \& Kochukhov, O.\ 2009, \aap, 499, 879
\bibitem[Shulyak et al.(2004)]{llm}Shulyak, D., Tsymbal, V., Ryabchikova, T., St\"utz\, Ch., \& Weiss, W. W. 2004, \aap, 428, 993
\bibitem[Thompson et al.(1978)]{1978csuf.book.....T} Thompson, G.~I., Nandy, K., Jamar, C., et al.\ 1978, The Science Research Council, U.K.
\bibitem[Tokovinin(1984)]{1984PAZh...10..293T} Tokovinin, A.~A.\ 1984, Pis'ma Astronomicheskii Zhurnal, 10, 293 
\bibitem[Tsymbal et al.(2003)]{2003IAUS..210P.E49T} Tsymbal, V., Lyashko, D., \& Weiss, W.~W.\ 2003, Modelling of Stellar Atmospheres, 210, 49P
\bibitem[van Leeuwen(2007)]{leeuwen}van Leeuwen, F. \ 2007, A\&A, 474, 653
\bibitem[Villemoes et al.(1992)]{1992PhyS...46...45V} Villemoes, P., Arnesen, A., F., Heijkenskj\'old, A, et al. \ 1992, Phys. Scripta, 46, 45  
\end{thebibliography}
\end{document}